\documentstyle[prl,aps,epsf,multicol]{revtex}
\tighten
\renewcommand{\narrowtext}
{\begin{multicols}{2}
\global\columnwidth20.5pc}
\renewcommand{\widetext}
{\end{multicols}\global\columnwidth42.5pc}
\begin{document} 
\draft 
\title{Critical Conductance of a Mesoscopic System: Interplay of the 
Spectral and Eigenfunction Correlations at the Metal-Insulator Transition}
\author{D. G. Polyakov$^*$}
\address{Institut f\"ur Theorie der
Kondensierten Materie, Universit\"at Karlsruhe, 76128 Karlsruhe,
Germany}
\maketitle 
\begin{abstract} 
We study the system-size dependence of the averaged critical
conductance $g(L)$ at the Anderson transition. We have: (i) related
the correction $\delta g(L)=g(\infty)-g(L)\propto L^{-y}$ to the
spectral correlations; (ii) expressed $\delta g(L)$ in terms of the
quantum return probability; (iii) argued that $y=\eta$ -- the critical
exponent of eigenfunction correlations. Experimental implications are
discussed.
\end{abstract} 
\narrowtext

Energy-level statistics and wavefunction correlations are two basic
issues in the theory of disordered electron systems at the Anderson
metal-insulator transition. So far, however, progress here has been
largely limited to scaling arguments in combination with numerical
simulations -- for lack of any parameter at the critical point. Our
interest in this problem owes much to the recent works
\cite{chalker96,chalker96jetp}, where a remarkably simple relation
between two fundamental characteristics of the critical system, the
spectral compressibility and the critical exponent of the
eigenfunction correlations, was {\it derived} analytically.

In the present paper we link the basic idea in
\cite{chalker96,chalker96jetp} with the problem of long-term kinetic
correlations at the metal-insulator transition. Specifically, we
address the question as to the mechanism of the system-size ($L$)
dependence of the ensemble-average critical conductance $g(L)$. In
accordance with the general principle of scale invariance at the
critical point, the critical conductance at the Anderson transition
\cite{lee85} converges at large $L$ to a scale-invariant value
$g(\infty)\sim 1$ (in units of $e^2/h$). What we are interested in is
thus the correction $\delta g(L)=g(\infty)-g(L)$, which breaks the
macroscopic scale invariance and behaves as \begin{equation} \delta
g(L)\sim (L_c/L)^y \end{equation} with a critical exponent $y>0$. The
microscopic length scale $L_c$ is defined by the condition $\delta
g(L_c)\sim 1$ and for the conventional Anderson transition in three
dimensions ($d$) coincides with the mean free path $l$. Most of our
discussion will be focused on the $2d$ case; namely, on the integer
quantum Hall (QH) critical point. The dissipative conductance at the
QH transition obeys Eq.\ (1) with the length $L_c$ given by $\ln
(L_c/l)\sim g^2(l)$; so that, in principle, the ``elementary block"
scale $L_c$ may be much larger than $l$. For concreteness, however, we
assume in what follows that $g(l)\sim 1$ and $L_c\sim l$. This is the
case, e.g., at the plateau transition in the lowest Landau level. We
seek a universal relation between the exponent $y$ and other critical
exponents, independent of the microscopic details.

Our main result is the following. We argue that \begin{equation}
\delta g(L)\propto\left[\int_\tau^{L^2/D}dtP(0,t)\right]^{-1}~,
\end{equation} where $P(0,t)$ is the quantum return probability in
time $t$, $\tau$ the scattering time, $D\propto L^{2-d}$ \cite{lee85}
the long wavelength diffusion coefficient (independent of $L$ in the
$2d$ case). The above integral is familiar from the weak-localization
theory \cite{lee85}: in the simplest case of time-reversal symmetry,
the finite-size correction to $g\gg 1$ in $2d$ is proportional to it
and grows with $L$ as $\ln (L/l)$. At the critical point the integral
diverges more strongly -- in a power-law manner. What we show is that
at the transition the scale-dependent correction to the conductance is
given by the {\it inverse} of the integral. It follows that the
exponent $y$ is not an independent one, but is equal to the critical
exponent of eigenfunction correlations $\eta$ \cite{chalker88}, which
controls behavior of $P(0,t)\propto t^{-1+\eta/d}$ at large $t$.

We start by defining two disorder-averaged two-particle correlators
\begin{eqnarray} P_\omega(r)={2\pi\over\rho}\left<\rho_\epsilon(0,{\bf
r})\rho_{\epsilon+\omega}({\bf r},0) \right>~,\\\nonumber
M_\omega(r)={2\pi\over\rho}\left<\rho_\epsilon(0,0)\rho_{\epsilon+\omega}
({\bf r},{\bf r}) \right>~, \end{eqnarray} where the one-particle
spectral function at energy $\epsilon$ reads $\rho_\epsilon({\bf
r},{\bf
r}')=\sum_\alpha\delta(\epsilon-\epsilon_\alpha)\psi_\alpha({\bf
r})\psi^*_\alpha({\bf r}')$ and $\alpha$ labels eigenstates in a
given realization of disorder. Throughout the paper we neglect the
$\epsilon$ dependence of disorder-averaged quantities, in particular
of the averaged density of states $\rho=\left<\rho_\epsilon\right>$,
where $\rho_\epsilon=L^{-d}\int d^d{\bf r}\rho_\epsilon({\bf r},{\bf
r})$, $L^d$ is the system volume. The normalization of the first
(``kinetic") correlator $\int d^d{\bf
r}P_\omega(r)=2\pi\delta(\omega)$ reflects particle number
conservation, while $\int d^d{\bf
r}M_\omega(r)=2\pi\Delta^{-1}(1+R_\omega)$ is expressed in terms of
the correlator of spectral fluctuations \begin{equation}
R_\omega={1\over\rho^2}\left<\delta\rho_\epsilon\delta\rho_{\epsilon+\omega}
\right>~.\end{equation} In the above, $\Delta=1/\rho L^d\to 0$ is the
mean level spacing.

We now write down the identity \begin{eqnarray}
{1\over\Delta}{\partial^2
R_\omega\over\partial\Lambda^2}&=&{\partial^2\over\partial\omega^2}\left<
\sum_\beta \left({\partial\omega_{\alpha\beta}\over\partial\Lambda}
\right)^2\delta (\omega_{\alpha\beta}+\omega)\right> \\\nonumber &+&
{\partial\over \partial\omega}\left<\sum_\beta {\partial^2
\omega_{\alpha\beta}\over\partial\Lambda^2}\delta(\omega_{\alpha\beta}
+\omega) \right>~, \end{eqnarray} where $\Lambda$ is an arbitrary
parameter characterizing the Hamitonian of the system $H(\Lambda)$ and
$\omega_{\alpha\beta}=\epsilon_\alpha-\epsilon_\beta$. We specify the
parameter $\Lambda$ by considering $\partial^2
R_\omega/\partial\Lambda^2$ a second-order response of the system to
the small ($\Lambda\to 0$) perturbation $\delta H(\Lambda)=\Lambda
e^{i{\bf kr}}$. Doing so we in fact introduce a ${\bf k}$-dependent
spectral correlator $R_\omega (k,\Lambda)$. Next we define its Fourier
transform ${\cal R}_\omega (r,\Lambda)=\int {d^d {\bf k}\over
(2\pi)^d}e^{-i{\bf kr}}R_\omega (k,\Lambda)$ and expand in powers of
$\Lambda$. To second order in $\Lambda$ [Eq.\ (5)] the expansion of
${\cal R}_\omega (r)$ readily yields \begin{eqnarray}
&&{1\over\Delta}{\partial^2 {\cal R}_\omega
(r)\over\partial\Lambda^2}=-{1\over\pi}{\partial^2 M_\omega
(r)\over\partial\omega^2}\\ &&+
{2\over\rho^2\Delta}{\partial\over\partial\omega}\int{d\omega'\over\omega'}
\left<\rho_\epsilon (0,{\bf r})\rho_{\epsilon+\omega'}({\bf
r},0)\left(\rho_{\epsilon+\omega+\omega'}-\rho_{\epsilon+\omega}\right)
\right>~.\nonumber \end{eqnarray} We decompose the three-particle
correlator in the second line of this equation according to the
following pattern \begin{eqnarray}
&&\rho\left<\delta\rho_{\epsilon_1}(0,{\bf
r})\delta\rho_{\epsilon_2}({\bf r},0)\delta\rho_{\epsilon_3}\right> \\
&&\simeq\left<\delta\rho_{\epsilon_1}(0,{\bf
r})\delta\rho_{\epsilon_2}({\bf
r},0)\right>\left(\left<\delta\rho_{\epsilon_1}\delta\rho_{\epsilon_3
}\right>+\left<\delta\rho_{\epsilon_2}\delta\rho_{\epsilon_3}\right>
\right)~.\nonumber \end{eqnarray} Note the extreme ``non-Gaussian"
character of the correlations: the irreducible three-particle average
is represented as a product of two pairwise correlators. The crucial
assumption behind this decoupling is that the correlations can be
separated into fast and slow (in energy space) pieces: namely, the
{\it spectral} correlations [Eq.\ (4)] decay rapidly on an energy
scale which vanishes in the thermodynamic limit, whereas the {\it
eigenfunction} correlations [Eq.\ (3)] decay slowly according to the
diffusion law. In effect, if one puts $r=0$, Eq.\ (7) constitutes
precisely what was incorporated in the Langevin description (and was
called a ``$t$-space decoupling") in \cite{chalker96,chalker96jetp}.
When substituted into Eq.\ (6), it gives, for the last term in the
r.h.s., \begin{equation} {1\over\pi\Delta}\int
{d\omega'\over\omega'}P_{\omega'}(r){\partial\over\partial\omega}
(R_{\omega+\omega'}-R_{\omega-\omega'})~. \end{equation}

The decomposition above is a peculiar property of the average (7) and
we proceed to discuss the conditions under which it is legitimate (the
conditions will equally apply to the method of
\cite{chalker96,chalker96jetp}). If $R_\omega$ would be characterized
by a single scale on which it falls off {\it sharply}, the separation
of the fast and slow variables would simply require that $P_\omega
(r)$ be a smooth function of $\omega$ on this scale. The spectral
correlations (in closed systems) indeed start to decay on the smallest
scale of $\Delta$; specifically, $R_\omega\propto\omega^{-2}$ at
$\Delta\alt |\omega|\alt E_c$ \cite{mehta}, where $E_c\sim g\Delta$ is
the Thouless energy, and continue to decay at larger
$|\omega|$. However, the argument that the energy-level correlations
are short ranged at $\Delta\to 0$ may be illusive, since $R_\omega$
falls off at $|\omega|\agt E_c$ in a power-law manner, which
necessitates a power counting. What really matters to the
justification of the above decoupling is {\it how} fast $R_\omega$
vanishes at large $|\omega|$. E.g., one sees that if $R_\omega$ decays
slower than $\omega^{-1}$, all frequencies in the range $|\omega'|\alt
|\omega|$ will give roughly the same contribution to the integral over
$\omega'$ in Eq.\ (8) -- in contrast to what the decoupling
implies. The sought conditions can thus be formulated as follows: (i)
$R_\omega$ falls off with increasing $|\omega|$ not slower than
$\omega^{-1}$; (ii) $P_\omega(0)$ does not decrease with $|\omega|$
faster than $\omega^{-1}$. Here and below we mean the frequency
dependence in the diffusive regime $E_c\alt |\omega| \alt\tau^{-1}$.

Clearly, the decomposition in terms of the pairwise correlators fails
in the insulating phase, where $P_\omega (r)$ is singular at
$\omega=0$ [see condition (ii)] even in the thermodynamic limit. So
the real question is about accuracy of the decoupling in a system with
finite $g$, where the condition on $P_\omega(r)$ is not severe but the
behavior of $R_\omega$ [see condition (i)] becomes crucial. It is
worthwhile to emphasize at this point that the validity of the
decoupling is not related directly to the expansion in powers of $1/g$
and the above conditions can be satisfied (at $\Delta/|\omega|\ll 1$)
even if $g\sim 1$. On the other hand, it is important that the
diagrammatic analysis \cite{chalker96} up to three-loop order does
conform to these conditions. Consider particular examples. The
frequency dependence of $R_\omega$ becomes steeper as the localization
effects get stronger and so does it with decreasing dimensionality. In
the $3d$ case, at $g(l)\gg 1$, $R_\omega$ decays too slowly -- as
$|\omega|^{-1/2}$ \cite{altshuler86} -- which means that the
decoupling certainly fails deep in the metallic phase. In the critical
regime of the metal-insulator transition, $R_\omega$ behaves as
$|\omega|^{-x}$ at $|\omega|\agt\Delta$ and how $x$ is related to
other critical exponents is still an open problem
\cite{aronov95}. However, the remarkable point to notice is that the
decoupling which fails in the metallic phase may work at the
transition, provided $x>1$. In $2d$, $R_\omega$ falls off at $g\gg 1$
as $|\omega|^{-1}$ \cite{kravtsov95}, which is a marginal case in
respect of the validity of the decoupling (meaning logarithmic
accuracy). The $\omega^{-1}$ behavior of $R_\omega$ at large $g$
implies that $x>1$ at the QH transition, which is one of the above
conditions. Now, we must also be concerned about how $P_\omega (r)$
behaves as $\omega\to 0$. At the transition $P_\omega (r)$ diverges at
small $\omega$, down to $|\omega|\sim\Delta$, as $|\omega|^{-\eta/d}$
\cite{chalker88}. To decouple the spectral and eigenfunction
correlations in $2d$, one needs $\eta < 2$ ($\eta=2$ corresponds to an
infinitely sparse fractal). At the QH critical point $\eta\simeq 0.4$
\cite{chalker88}). We conclude that the decoupling is valid at the QH
transition \cite{aside1}.

From now on we specialize to the QH transition. The net result of the
decoupling (7) is that the expansion of ${\cal R}_\omega (r)$ in
$\Lambda$ for any given $\omega$ and $r$ yields just $M_\omega(r)$ and
$P_\omega(r)$. It is convenient to transform to the time
representation -- introducing $M(r,t)$ and $P(r,t)$ -- where the
relation between the two functions assumes a particularly compact
form: \begin{eqnarray} tM(r,t)-2K\int_0^tdt'P(r,t')={1\over
2t}{\partial^2{\cal K}(r,t)\over\partial\Lambda^2} ~.\end{eqnarray}
Here we have also introduced the dimensionless Fourier transform
$K(t)=\Delta^{-1}\int d\omega e^{-i\omega t}R_\omega$ and a similar
transform ${\cal K}(r,t)$ for ${\cal R}_\omega (r)$. It is worth
emphasizing that the validity of the decoupling (the condition (i)
above) requires $K(t)$ to be constant in the thermodynamic limit
\cite{aside2}. For our purposes, we thus may write $R_\omega=K\delta
(\omega/\Delta)+(\Delta/\Sigma) F(\omega/\Sigma)$, where
$\Sigma=\tau^{-1}$, $F(x)$ is a dimensionless function of order
unity. It is the last term that is crucial to the analysis of the
finite-size scaling. Clearly, it violates the condition (i);
nevertheless, Eq.\ (9) is useful if we are interested only in the
critical exponents -- not the exact (nonuniversal) shape of
$F(x)$. Indeed, at $r=0$ we recover the relation
\cite{chalker96jetp}\begin{equation} K=\lim_{t/\tau\to\infty}{1\over
2}tP(0,t)\left(\int_0^tdt'P(0,t')\right)^{-1}~, \end{equation} which,
taken at finite $t/\tau$, allows to estimate the small $t$-dependent
term in $K(t)$. In this way we find that $K(\infty)-K(t)\sim
(\tau/t)^{\eta/2}$ (i.e., $F(x)$ diverges as $x^{-1+\eta/2}$ at $x\to
0$). Note that in going to the last step we have neglected the
parametric correlations in the r.h.s.\ of Eq.\ (7) (while in
\cite{chalker96} they were omitted by construction), which needs
explanation. We first rewrite $\partial^2{\cal
K}(r,t)/\partial\Lambda^2$ at $r=0$ as the response $\partial^2
K(t)/\partial u^2$ to the application of a white-noise random
potential $W({\bf r})$ with the correlator $\left<WW\right>=u^2\delta
({\bf r})$. This additional random potential merely renormalizes the
scattering rate. It follows that \begin{equation} {\partial^2{\cal
K}(0,t)\over\partial\Lambda^2}=4\pi\rho{\partial
K(t)\over\partial\Sigma}~. \end{equation} (recall that, for
simplicity, we keep $\rho$ independent of $\Sigma$, even at $g\sim
1$). Hence the starting idea in \cite{chalker96}, about the
statistical equivalence of all members of the ensemble of
Hamiltonians, amounts to ignoring the dependence of $K(t)$ on
$\Sigma$. One can easily estimate the omitted contribution to $K(t)$:
it is $\sim(\tau/t)^{1+\eta}$, which is indeed smaller than the
leading correction obtained above. Now we can argue in the same way to
make sure that if $t$ is large enough, the parametric correlations can
be neglected at finite $r$ as well. The large-$t$ limit means
$t\gg\max\{\tau, r^2/D\}$, so that as $r$ increases it needs more time
for the parametric correlations to be forgotten. It is also
instructive to see how in $2d$ the first terms in the expansion of
$M$, $P$, and $K$ in powers of $g^{-1}$ cancel in Eq.\ (9). Note that
$M(r,t)$ and $P(r,t)$ can be split into short and long ranged (in real
space) parts. Specifically, $M(r,t)\simeq 2\pi\rho\beta (0,t) +
f^2(r)p(0,t) + (2\pi\rho)^{-1}\int_0^t dt' p(r,t')p(r,t-t')$,
$P(r,t)\simeq 2\pi\rho\beta (r,t) + p(r,t)$, and $K\simeq (4\pi
g)^{-1}$. Here $\beta (r,t)$ describes ballistic motion at $t\alt\tau$
and vanishes at larger $t$, $\int dt \beta(r,t)=f^2(r)$,
$f(r)=e^{-r/2l}J_0(k_Fr)$ is the Friedel function ($k_F$ is the Fermi
momentum), and $p(r,t)$ is the retarded Green function of the
classical diffusion equation. For concreteness, we mean a system with
completely broken time-reversal symmetry. Since $K\sim g^{-1}$, we
need next order terms in $M(r,t)$ as compared to $P(r,t)$ -- hence the
two-diffuson term above. One can see that in the limit $t\gg\max
\{\tau,r^2/D\}$ all the terms cancel out in the l.h.s.\ of Eq.\
(9). Despite the simplicity, it is a useful check.

Now we write the correlators in the scaling form which they assume at
the QH transition: $M(r,t)={1\over Dt}{\rm M}({r^2\over Dt})$,
$P(r,t)={1\over Dt}{\rm P}({r^2\over Dt})$, where both ${\rm M}(x)$
and ${\rm P}(x)$ behave as $x^{-\eta/2}$ at $x\ll 1$ with $\eta\simeq
0.4$ \cite{chalker88}. The hydrodynamic form of the scaling functions
implies that $r\gg\lambda$ and $t\gg\tau$, where $\lambda$ is the
magnetic length and $D\sim\rho^{-1}\sim\lambda^2\tau^{-1}$ (for
definiteness we mean the case of short-range disorder in the lowest
Landau level). Clearly, within the scaling approach the average
conductance is scale independent. Indeed, whatever the function ${\rm
P}(x)$ is, the velocity-velocity correlator $G_v(t)=\int d^2{\bf
r}r^2{\partial^2 P(r,t)\over\partial t^2}$ is proportional, within the
scaling description, to $\delta (t)$. Hence, it is corrections to the
scaling that determine the asymptotic behavior of $G_v(t)$ at
$t\gg\tau$. We expect that the quantum kinetic correlations decay at
the transition in a power-law manner: \begin{equation} G_v(t)\sim
{D\over \tau}\left({\tau\over t}\right)^{1+{y\over 2}}
~. \end{equation} 

We now appeal to Eq.\ (9) in order to find $y$. The essential idea is
that the tail in $G_v(t)$ originates from a power-law (in $\tau/t\ll
1$) correction to the limiting value of $K(t)$ at
$t\to\infty$. Indeed, at the critical point $K(t)$ at $\Delta\to 0$ is
parameterized as a function of the single variable $t/\tau$. It
follows that the $r$-dependent correlators written in the scaling form
as functions of $r^2/Dt$ satisfy Eq.\ (9) only if $K(t)$ is strictly
constant. Hence, taking into account $t$-dependent terms in $K(t)$
will generate corrections to the scaling. As noted above, these yield
the sought tail in $G_v(t)$. We thus expand $K(t)$ in $\tau/t$ and
substitute the leading term in the expansion into Eq.\
(9). Multiplying the equation by $r^2$ and doing the integrals over
$r$, we obtain the contribution to $G_v(t)$ which is due to the
dispersion of $K(t)$ -- it is given by Eq.\ (12) with
\begin{equation}y=\eta~. \end{equation} The exponent $y$ defined by
Eq.\ (12) governs behavior of the average conductance $g(L)$ in a
finite sample of size $L\times L$. Specifically, the conductance at
$L\to\infty$ is given by the Kubo formula for the conductivity
$g(\infty)={\pi\over 2}\rho\int_0^\infty dt G_v(t)$, and cutting
the integral over $t$ at $t\sim L^2/D$ yields $g(L)$. Since we
associate the long-term tail in $G_v(t)$ with the decay of the
difference $K(\infty)-K(t)$, which in turn is given by the behavior of
$P(0,t)$ [Eq.\ (10)], we finally arrive at Eq.\ (2).

A cautionary remark is in order at this point. We assumed that the
corrections to $P(r,t)$ and $M(r,t)$ do not cancel each other in Eq.\
(9) exactly unless other irrelevant scaling fields are taken into
account -- since, apart from the common exponent $y$, the corrections
depend on microscopic details (say the correlation properties of
disorder). Eq.\ (9) thus tells us that there {\it is} a contribution
to $G_v(t)$ which scales with $y=\eta$. However, because of the
parametric correlations in the r.h.s., we cannot rule out solely on
the basis of this equation the possibility that there exists another
contribution with smaller $y$ (a ``less irrelevant" scaling field),
not at all related to the behavior of $K(t)$. Therefore, to be
precise, what we have been able to prove is the {\it inequality}
$y\leq\eta$. At this point we might seek support from numerical
data. These suggest strongly that the sign $\leq$ above is in fact
``equals": indeed, the values of $y\simeq 0.4-0.5$
\cite{huckestein95,wang96} and $\eta\simeq 0.4$
\cite{chalker88,huckestein95} obtained in the numerical simulations
are fairly close to each other.

So far we have dealt with the scaling of the {\it average} conductance
$g(L)$. Of what significance could it be in real experiments in view
of the fact that at the critical point the mesoscopic fluctuations are
very strong? To extract the scale dependent $\delta g$ at the QH
transition, the dephasing length $L_\phi\sim (D/T)^{1/2}$ (or,
interchangeably, $L_\omega\sim (D/\omega)^{1/2}$ in ac measurements)
should be much smaller than the sample size -- the {\it measured}
conductance then will be a self-averaging well-defined quantity
$g^{\rm eff}(L_\phi)$ (independent of $L$ in $2d$) with a small
temperature dependent correction $\delta g^{\rm eff}(L_\phi)\sim
(\lambda/L_\phi)^y$. In this respect, the experiment with the QH metal
looks much like the measurement of the localization corrections in a
good metal. Note two points relevant to the experiment. First, the
average (``coherent") conductance $g(\infty)$ (at
$L/\lambda\to\infty$) and the measured at finite $T$ (``effective")
conductance $g^{\rm eff}(\infty)$ (at both $L_\phi/\lambda\to\infty$
and $L/L_\phi\to\infty$) are by no means the same. One way to look at
the problem is to divide the system at the critical point into blocks
of size $L_\phi$ and couple them to each other ``incoherently", thus
representing the system as a classical resistance network which obeys
the Ohm's law locally. Since the quantum interference is preserved
inside the blocks, the elementary conductances $g_i$ of this network
will fluctuate strongly around mean $g(L_\phi)$. One may now consider
expanding $g^{\rm eff}-g$ in series in the cumulants of $g_i$
\cite{lanlif} -- the first term in the expansion is given by
$-(\left<g_i^2\right>-\left<g_i\right>^2)/2\left<g_i\right>$. Since
there is no small parameter in this expansion, it follows immediately
that $|g^{\rm eff}(\infty)-g(\infty)|\sim 1$. What is important to us,
however, is that the scaling of the corrections to either of the
quantities is governed by the same exponent $y$. The difference
between the measured and averaged conductances is nonetheless crucial
when referring to the concept of the universal $g$ at the quantum
phase transition (in this respect the above arguments complement those
in \cite{dample97}). Second, the applicability of the non-interacting
model considered in the paper to the experiment at the integer QH
critical point requires comment. There exists a long-range
contribution to the velocity-velocity correlator $G_v(t)$ associated
with electron-electron interactions: as is well known, at large $g$ in
$2d$ it decays as $t^{-1}$ \cite{lee85} -- i.e., it scales similarly
to the weak-localization contribution. We have shown that the $t^{-1}$
tail which is due to the localization effects is transformed at the
integer QH transition into $t^{-1-\eta/2}$. The question as to what
extent the Coulomb interaction affects this behavior of $G_v(t)$
requires further study.

In conclusion, we have studied how the critical conductance $g(L)$ at
the metal-insulator transition behaves as a function of the system
size $L$. We relate the size dependent correction $\delta g(L)\sim
(l/L)^y$ to the quantum return probability and argue that the exponent
$y$ is the same as $\eta$ -- the critical exponent of eigenfunction
correlations. 

I thank J. Chalker and A. Mirlin for numerous interesting
discussions. I am especially grateful to M. Janssen who participated
in this work at its early stage. The hospitality of the Theoretical
Physics Institute at the University of Cologne, where part of this
work was done, is gratefully acknowledged. The work was supported by
the Deutsche Forschungsgemeinschaft through SFB 195 and by the
German-Israeli Foundation.

\end{multicols}

\begin{references}
\bibitem{aside1} The decoupling obviously fails at small wavevectors
$qL\alt 1$ -- since the kinetic propagator $P_{\omega q}$ becomes
singular at $q\to 0$ -- but this range of $q$ vanishes at $\Delta\to
0$.
\bibitem{aside2} This resolves the problem posed by the introduction
and comparative analysis of two different decoupling schemes in Ref.\
[1] (product vs.\ convolution of $K(t)$ and $P(0,t)$). If $K(t)$ is
constant, one can choose either scheme; otherwise, neither of them is
legitimate.
\end{references}
\end{document}